\documentclass[a4paper,11pt]{article}
\usepackage{pos}
\usepackage[utf8]{inputenc}

\usepackage{amsmath,amssymb,mathtools,booktabs}
\usepackage{graphicx}
\usepackage{float}
\usepackage{microtype}
\usepackage{hyperref}
\hypersetup{colorlinks=true, linkcolor=blue, citecolor=blue, urlcolor=blue}

\title{B-Meson Anomalies: Effective Field Theory Meets Machine Learning}
\ShortTitle{EFT Meets ML for B Anomalies}

\author*[a,b]{A.~Mir}
\author[b,c,d]{J.~Alda}
\author[a,b]{S.~Pe\~naranda}

\affiliation[a]{Departamento de F\'isica Te\'orica, Universidad de Zaragoza, Pedro Cerbuna 12, E-50009 Zaragoza, Spain}
\affiliation[b]{CAPA, Centro de Astropart\'iculas y F\'isica de Altas Energ\'ias, Universidad de Zaragoza, E-50009 Zaragoza, Spain}
\affiliation[c]{Dipartimento di Fisica ed Astronomia ``Galileo Galilei'', Universit\`a di Padova, via F. Marzolo 8, 35131 Padova, Italy}
\affiliation[d]{INFN, Sezione di Padova, via F. Marzolo 8, 35131 Padova, Italy}

\emailAdd{amir@unizar.es}
\emailAdd{jorge.alda@pd.infn.it}
\emailAdd{siannah@unizar.es}

\abstract{
Discrepancies between experimental measurements and Standard Model predictions in $B$-meson decays, especially in lepton flavor universality ratios like $R_{D^{(*)}}$, $R_{J/\psi}$ and branching ratios for processes like $B\to K^+\nu\bar\nu$, suggest possible new physics (NP). In this study, we use an effective field theory framework, assuming NP effects only affect a single generation in the interaction basis, leading to non-universal mixing when rotating to the mass basis. We perform a global fit to the current experimental data, exploring three scenarios characterized by different mixing patterns and constraints. Our analysis finds that the best fit involves mixing between the second and third quark generations, with no lepton sector mixing and independent coefficients for singlet and triplet four-fermion operators. To accurately capture the non-Gaussian nature of the resulting parameter distributions, we use a Machine Learning-based Monte Carlo algorithm, enabling the generation of representative samples that reflect the true underlying distributions. This work highlights the valuable role of Machine Learning in accurately modeling complex parameter distributions in particle physics analyses.
}

\FullConference{The European Physical Society Conference on High Energy Physics (EPS-HEP2025)\\ 7-11 July 2025\\Marseille, France}

\note{This work is partially supported by Spanish MINECO/FEDER Grants No. PGC2022-126078NBC21 funded by MCIN/AEI/10.13039/ 501100011033 and “ERDF A way of making Europe” and Grant DGA-FSE Grant 2020-E21-17R Aragón Government and the European Union- NextGenerationEU Recovery and Resilience Program on Astrofísica y Física de Altas Energías CEFCA-CAPA13ITAINNOVA.}

\begin{document}

\pagenumbering{gobble}
\maketitle
\thispagestyle{empty}
\clearpage
\pagenumbering{arabic}
\setcounter{page}{1}
\pagestyle{plain}
\thispagestyle{plain}

Semileptonic $B$ meson decays provide sensitive tests of the Standard Model (SM) through lepton flavour universality (LFU) ratios and rare modes with neutrinos in the final state, $b\to c \ell \nu$ and $b\to s \,\ell^+ \ell^-$ transitions. Persistent differences between measurements and precise SM predictions could be a sign for physics beyond the SM. Values of the SM predictions and experimental results for $R_D, R_{D^*}, R_{J/\psi}$ and $\mathrm{BR}(B^+\to K^+\nu\bar{\nu})$ are collected in Table~\ref{tab:inputs} and set the scale of the effects discussed here. The effective field theory (EFT) analysis provides a model-independent study of these effects and treats charged and neutral current sectors on equal footing and evaluates their correlations under controlled flavour assumptions. The aim is a minimal description of left-handed semileptonic modifications compatible with present data, together with a clear account of how information from diverse observables constrains the parameter space. The influence of several effective operators on numerous observables is mediated through their connection via the Wilson coefficients. In order to understand the impact of experimental results on flavour physics observables, it is necessary to perform a global fit to the available experimental data. We performed a global fit in previous articles~\cite{Alda:2022ML,Alda:2020okk,Alda:2412.15830}, where an extensive list of references is included. In~\cite{Alda:2412.15830} the $b\to c \tau \bar{\nu}$ and the $B\to K^{(*)} \nu \bar{\nu}$ decays are analyzed, providing the best fit predictions for $R_{D^{*}}$ and $R_{J/\psi}$ observables and for the $B^+\to K^+\nu\bar{\nu}$ and $B^0\to K^{*0}\nu\bar{\nu}$ branching ratios. 

\begin{table}[H]
\centering
\small
\begin{tabular}{@{}lcc@{}}
\toprule
Observable & SM & Experiment \\
\midrule
$R_D$ & $0.299 \pm 0.004$ & $0.342 \pm 0.026$ \\
$R_{D^*}$ & $0.257 \pm 0.005$ & $0.287 \pm 0.012$ \\
$R_{J/\psi}$ & $0.258 \pm 0.004$ & $0.52 \pm 0.20$ \\
$\mathrm{BR}(B^+\to K^+\nu\bar\nu)$ & $(0.429 \pm 0.021)\times10^{-5}$ & $(2.3 \pm 0.7)\times10^{-5}$ \\
\bottomrule
\end{tabular}
\caption{Representative inputs (SM vs.\ experiment) used in the analysis~\cite{HFLAV:2022esi}.}
\label{tab:inputs}
\end{table}

New-physics (NP) effects are encoded through two left-handed semileptonic structures organised as weak-isospin singlet and triplet, with flavour projectors selecting a dominant component in lepton and quark spaces:
\begin{equation}
\mathcal{L}_{\rm NP}=\frac{\lambda_\ell^{\,ij}\lambda_q^{\,kl}}{\Lambda^2}\left[
C_1(\bar\ell_i\gamma_\mu \ell_j)(\bar q_k\gamma^\mu q_l)+
C_3(\bar\ell_i\gamma_\mu \tau^I\ell_j)(\bar q_k\gamma^\mu \tau^I q_l)\right],
\end{equation}
where $\lambda^\ell$ and $\lambda^q$ are the matrices that determine the flavour structure of the NP interactions. The projectors are hermitian and idempotent with ${\rm tr}\,\lambda=1$, parametrised by two complex numbers, $\alpha^{\ell,q}$ and $\beta^{\ell,q}$, as
\begin{equation}
\lambda_{\ell,q}=\frac{1}{1+|\alpha_{\ell,q}|^2+|\beta_{\ell,q}|^2}
\begin{pmatrix}
|\alpha_{\ell,q}|^2 & \alpha_{\ell,q}\overline{\beta}_{\ell,q} & \alpha_{\ell,q}\\
\overline{\alpha}_{\ell,q}\beta_{\ell,q} & |\beta_{\ell,q}|^2 & \beta_{\ell,q}\\
\overline{\alpha}_{\ell,q} & \overline{\beta}_{\ell,q} & 1
\end{pmatrix}.
\end{equation}

We examine three benchmark hypotheses: Scenario I with $C_1=C_3$ and quark mixing restricted to the second and third generations, no lepton mixing; Scenario II with $C_1=C_3$ and generic mixing in both sectors; Scenario III with $C_1\neq C_3$, 2--3 quark mixing only, no lepton mixing ($\alpha^\ell = \beta^\ell = 0$). In this setup, LFU ratios in $b\to c\ell\nu$ and the neutrino modes $B\to K^{(*)}\nu\bar\nu$ probe complementary combinations of $(C_1,C_3)$ along components selected by $\lambda_{\ell,q}$.

The fit includes all observables implemented by \texttt{smelli} version 2.3.3~\cite{Aebischer:2018iyb} plus the observable $R_{J/\psi}$ that we implemented using the form factors in~\cite{Harrison:2020gvo}. This choice ensures that charged and neutral current observables are fitted coherently since the coefficients $C_1$ and $C_3$ affect both types of currents, constrains the flavour mixing introduced by the quark projector $\lambda_q$ after rotating to the mass basis, and prevents improvements in one sector from reappearing as tensions elsewhere. While LFU ratios in $b\to c\ell\nu$ and the neutrino modes $B\to K^{(*)}\nu\bar\nu$ provide the main pull on $(C_1,C_3)$ under the hypotheses above, null tests, electroweak precision tests, CKM constraints and light-lepton channels contribute to breaking approximate degeneracies, profiling nuisances and validating the stability of the preferred region.

The analysis combines two complementary procedures. First, we determine the optimal parameters $(C_1, C_3, \beta_q)$ by minimizing the $\chi^2$ of the difference between experimental and theoretical values of the observables. This global fitting process provides a reference set of confidence intervals. Second, we construct a regression tree algorithm of the exact log--likelihood to accelerate high resolution mapping and uncertainty propagation.

We generate $10{,}500$ parameter points and evaluate the exact log--likelihood at each point. The sample is split evenly into training (84\%) and validation (16\%) sets. We use a gradient boosted ensemble of shallow decision trees (XGBoost) with learning rate $0.03$ and early stopping after $5$ rounds; training converges after approximately $776$ boosting iterations. On the validation set the emulator reproduces the targets with high linear correlation (Pearson coefficient $r\simeq 0.96$), and its level sets coincide with those of the exact likelihood within the plotting resolution used in the figures.

This tree based approach is advantageous here for three reasons aligned with our use case: (i) The confidence intervals are non Gaussian and can develop curved ridges in $(C_1,C_3,\beta_q)$; boosted trees approximate such structure without assuming global smoothness. (ii) Mixed parameter scales and weakly constrained combinations are handled robustly by tree partitions without preconditioning. (iii) Once trained, evaluations are fast and enable dense sampling to draw $1\sigma$-$2\sigma$ contours and to propagate uncertainties to the observable planes. For diagnostics and interpretability, we also compute SHAP (Shapley Additive Explanations) values on the trained tree ensemble, which quantify the contribution of each parameter $(C_1,C_3,\beta_q)$ to the emulator output both globally and point by point.

 Other Machine Learning algorithms, such as neural networks, can emulate the likelihood as well, but in this low dimensional, non Gaussian setting they typically require larger training samples and stronger regularisation to avoid over-smoothing narrow ridges. Regression trees achieve the required accuracy with modest tuning and provide stable performance under reseeding.

\begin{table}
\centering
\small
\renewcommand{\arraystretch}{1.15}
\begin{tabular}{@{}lccc@{}}
\toprule
 & \textbf{Scenario I} & \textbf{Scenario II} & \textbf{Scenario III} \\
\midrule
$C_1$ & $-0.11^{+0.03}_{-0.04}$ & $-0.12 \pm 0.03$ & $-0.205 \pm 0.015$ \\
$C_3$ & $-0.11^{+0.03}_{-0.04}$ & $-0.12 \pm 0.03$ & $-0.12^{+0.02}_{-0.01}$ \\
$\alpha^\ell$ & --- & $0.0 \pm 0.07$ & --- \\
$\beta^\ell$ & $0.00 \pm 0.02$ & $0.000 \pm 0.014$ & --- \\
$\alpha^q$ & --- & $-0.076$ & --- \\
$\beta^q$ & $0.78^{+1.22}_{-0.36}$ & $0.85^{+1.05}_{-0.60}$ & $0.64^{+1.36}_{-0.24}$ \\
\midrule
Pull & $5.71\,\sigma$ & $5.51\,\sigma$ & $6.25\,\sigma$ \\
$\Delta\chi^2_{\rm SM}$ & $39.8$ & $43.12$ & $46.66$ \\
$p$-value & $1.2\times10^{-8}$ & $3.5\times10^{-8}$ & $4.1\times10^{-10}$ \\
\bottomrule
\end{tabular}
\caption{Summary of best--fit parameters and goodness--of--fit for the three benchmark scenarios. Uncertainties are one--standard--deviation intervals.}
\label{tab:fit_results}
\end{table}
Table~\ref{tab:fit_results} collects the best--fit results for the three benchmarks; Scenario~III exhibits the largest pull and is therefore the preferred case, which is the one studied in the remainder of this proceedings.

\begin{figure}[t]
\centering
\begin{minipage}{0.46\textwidth}\centering
\includegraphics[width=\linewidth]{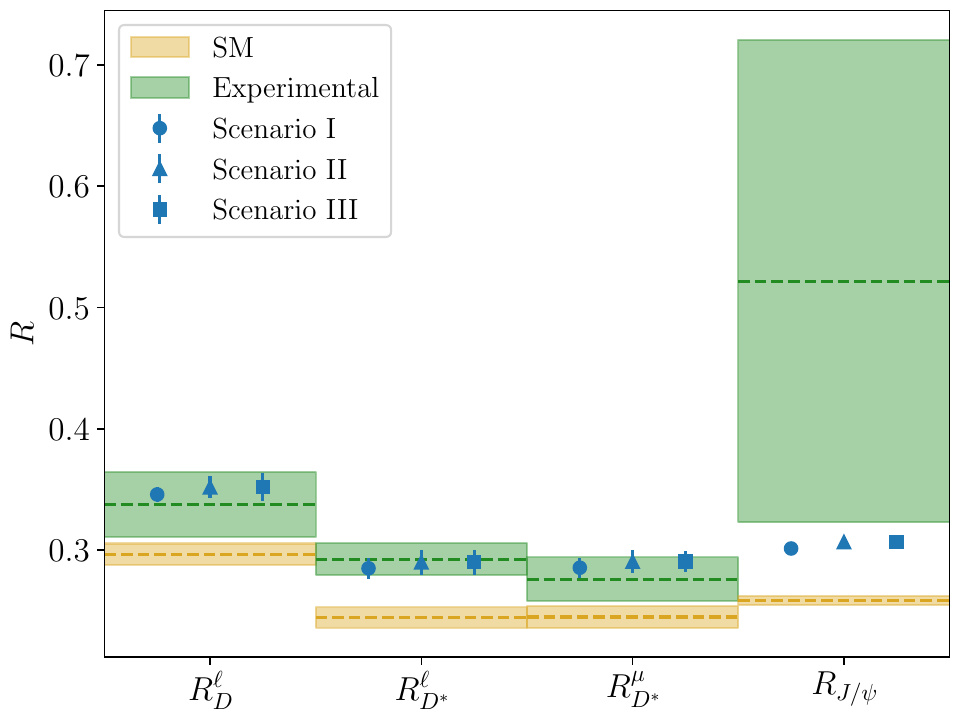}
\end{minipage}\hfill
\begin{minipage}{0.46\textwidth}\centering
\includegraphics[width=\linewidth]{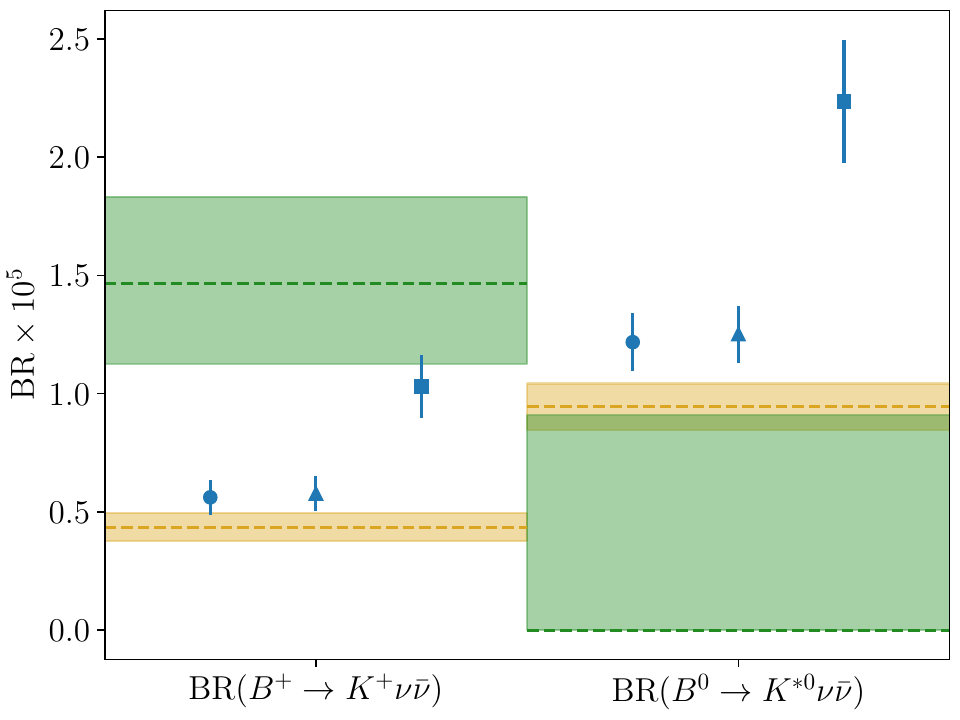}
\end{minipage}
\caption{Scenario III: selected LFU ratios in $b\to c\ell\nu$ (left) and the neutrino-mode branching fractions (right) in the preferred region and across scenarios.}
\label{fig:obs}
\end{figure}

Figure~\ref{fig:obs} presents predictions in the preferred region of Scenario III for the key observables used to confront the EFT hypotheses. For completeness, results for Scenarios I and II are also included. The left panel compiles LFU ratios from $b\to c\ell\nu$, while the right panel displays $\mathrm{BR}(B^+\to K^+\nu\bar\nu)$ and $\mathrm{BR}(B^0\to K^{*0}\nu\bar\nu)$. Enforcing $C_1=C_3$ over-correlates charged and neutral current sectors and hinders a simultaneous description. Relaxing this relation decouples the combinations of $(C_1,C_3)$ probed by the two classes of observables, yielding a coherent fit without inducing tension in channels that are SM-like at the order considered.

\begin{figure}[t]
\centering
\begin{minipage}{0.32\textwidth}\centering
\includegraphics[width=\linewidth]{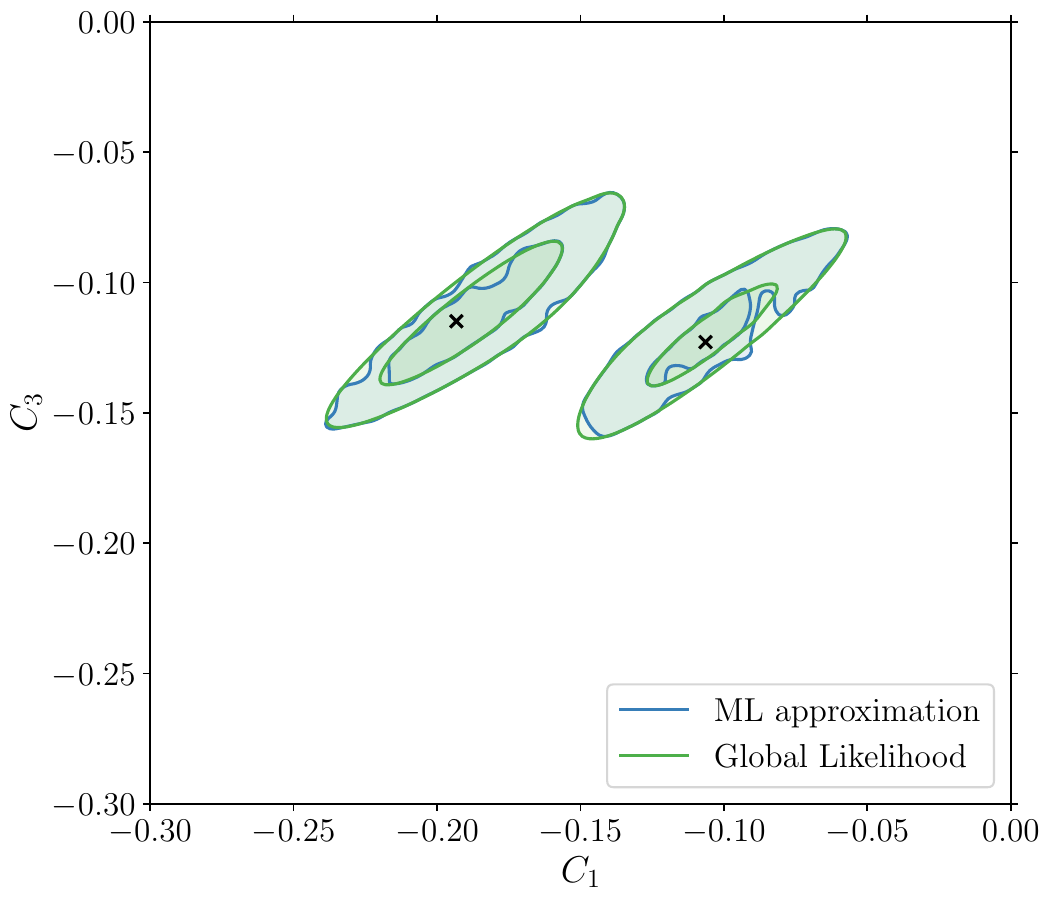}
\end{minipage}\hfill
\begin{minipage}{0.32\textwidth}\centering
\includegraphics[width=\linewidth]{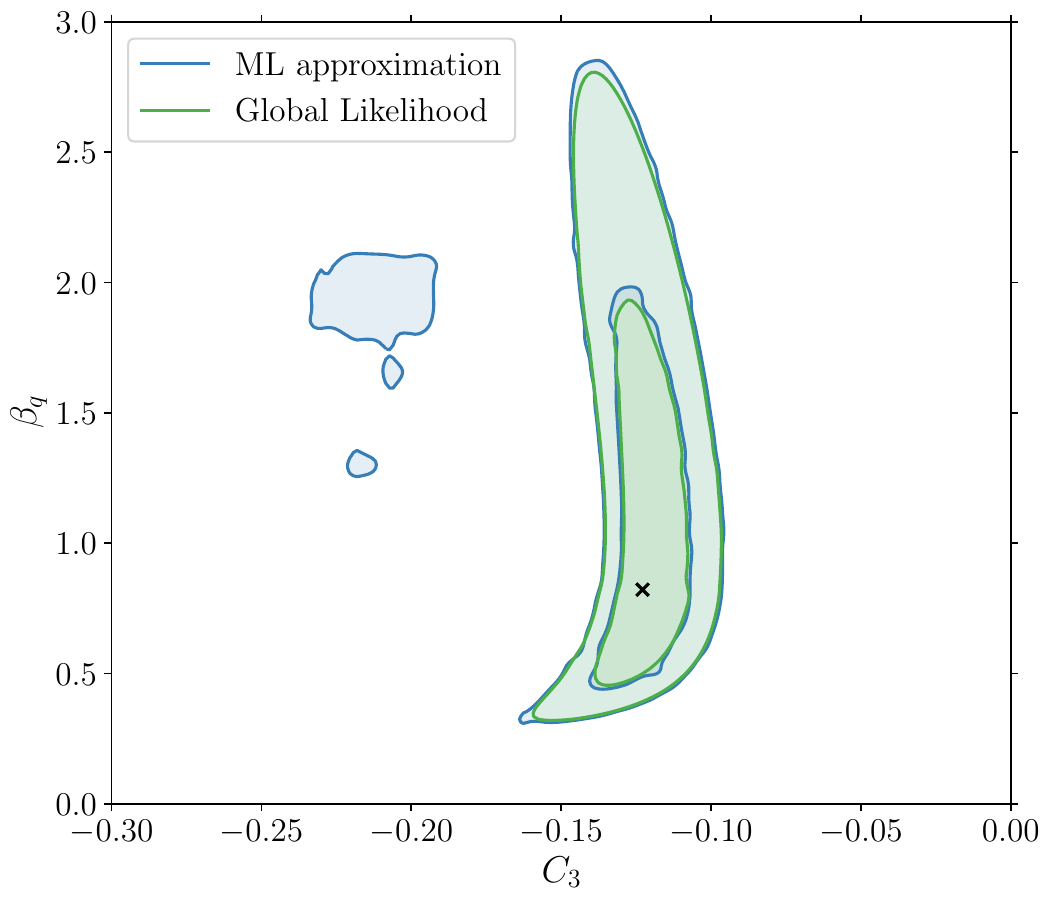}
\end{minipage}\hfill
\begin{minipage}{0.32\textwidth}\centering
\includegraphics[width=\linewidth]{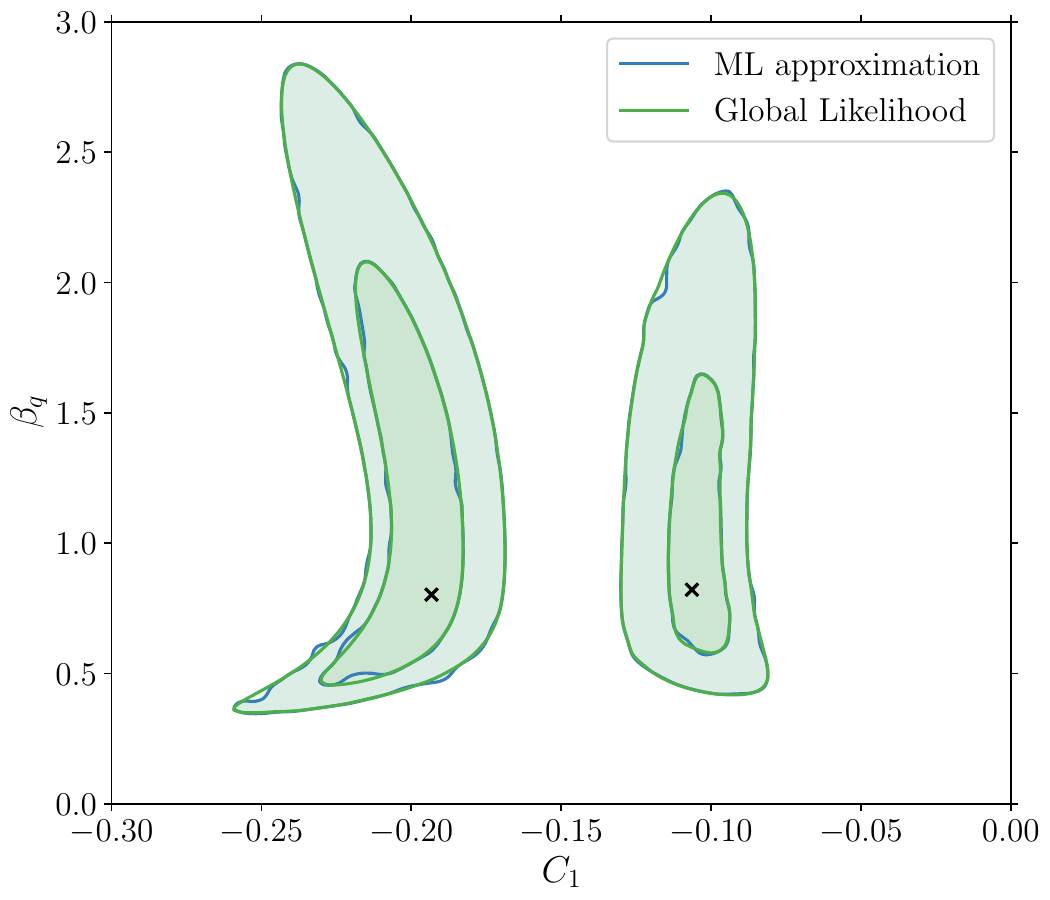}
\end{minipage}
\caption{Scenario III: two--parameter likelihood slices in $(C_1,C_3)$, $(C_3,\beta_q)$ and $(C_1,\beta_q)$. Regression--tree emulator and exact contours agree across the three planes.}
\label{fig:like}
\end{figure}

The comparison between likelihood structure of the fit and the Machine Learning algorithm prediction is shown in Figure~\ref{fig:like} in the planes $(C_1,C_3)$, $(C_3,\beta_q)$ and $(C_1,\beta_q)$ for Scenario III, showing a high degree of agreement. Therefore, we can conclude that ML reproduces correctly the general features of the fit and it is a suitable strategy for this kind of analysis. 

\begin{figure}[t]
\centering
\begin{minipage}{0.48\textwidth}\centering
\includegraphics[width=\linewidth]{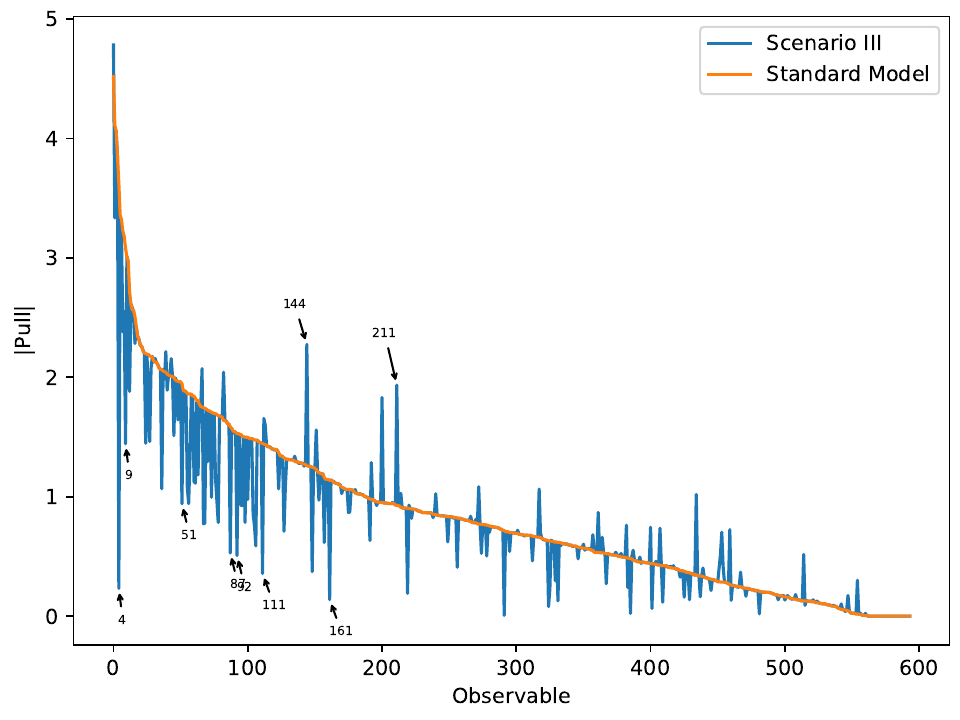}
\end{minipage}\hfill
\begin{minipage}{0.48\textwidth}\centering
\includegraphics[width=\linewidth]{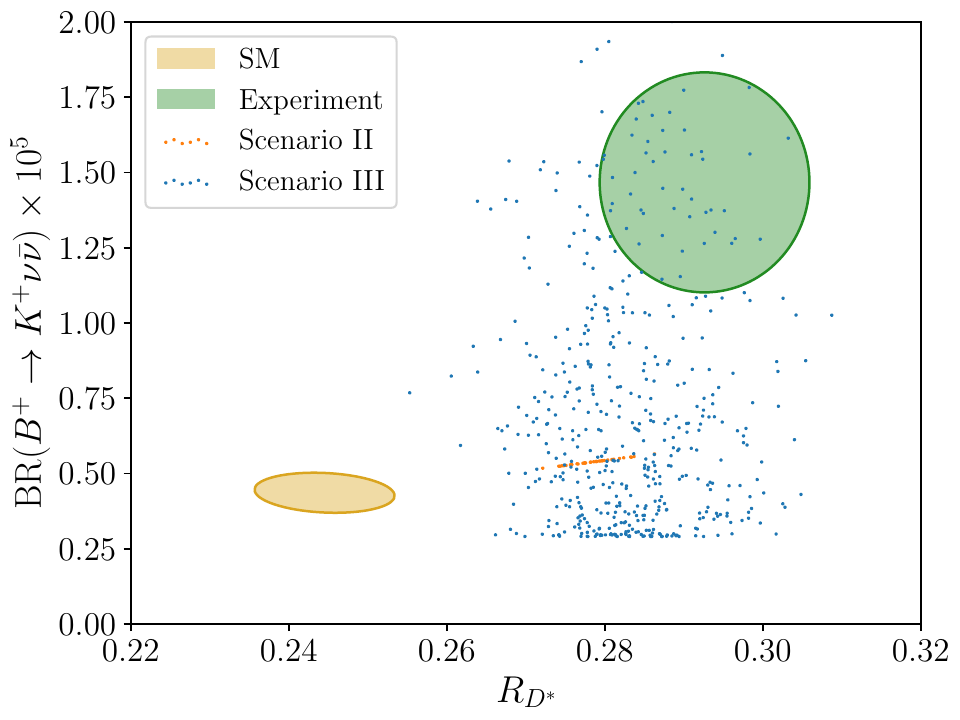}
\end{minipage}
\caption{Scenario III: left, pulls for the SM and the preferred configuration; right, samples in $(R_{D^*},\mathrm{BR}(B^+\to K^+\nu\bar\nu))$ showing that the trend present for $C_1=C_3$ vanishes for independent $(C_1,C_3)$.}
\label{fig:pulls-corr}
\end{figure}

A compact global view is provided in Figure~\ref{fig:pulls-corr}. The left panel compares the pull of each observable with respect to its experimental value in the SM (orange line) and in Scenario III (blue line). The numbers attached to some entries are indices of observables used in the full paper’s appendix table (not reproduced here); they are shown only to identify which observables differ by more than $1\,\sigma$. The highlighted observables are all related to the $B$ meson anomalies: $R_{D^*}^\ell$ (index 4) and $R_D^\ell$ (index 92) improve by $3.3\,\sigma$ and $1\,\sigma$, respectively; $\mathrm{BR}(B^+\!\to\!K^+\nu\bar{\nu})$ (index 9) improves by $1.7\,\sigma$ and $\mathrm{BR}(B^+\!\to\!K^{*+}\nu\bar{\nu})$ (index 161) by $1\,\sigma$, while $\mathrm{BR}(B^0\!\to\!K^{*0}\nu\bar{\nu})$ (index 144) becomes worse by $1\,\sigma$. For $b\to s\mu^+\mu^-$ at high $q^2$, $\mathrm{BR}(B^\pm\!\to\!K^\pm \mu^+\mu^-)$ in the $[16,17]\,\mathrm{GeV}^2$ (index 51) and $[15,22]\,\mathrm{GeV}^2$ (index 87) bins, as well as $\mathrm{BR}(B_s^\pm\!\to\!\phi^\pm \mu^+\mu^-)$ in the $[15,19]\,\mathrm{GeV}^2$ bin (index 111), improve by about $1\,\sigma$, whereas $\mathrm{BR}(B^\pm\!\to\!K^\pm \mu^+\mu^-)$ in the $[17,18]\,\mathrm{GeV}^2$ bin (index 211) worsens by about $1\,\sigma$. Within a setup with only left-handed effective operators it is not possible to accommodate simultaneously $B^+\!\to\!K^+\nu\bar{\nu}$ and $B^0\!\to\!K^{*0}\nu\bar{\nu}$ as shown in Figure~\ref{fig:obs}, which explains the deterioration observed for index 144. The right panel displays samples in $(R_{D^*},\mathrm{BR}(B^+\to K^+\nu\bar\nu))$: the tight positive trend that appears when $C_1=C_3$ disappears once the singlet and triplet coefficients are allowed to vary independently.

Summarising, an EFT with two left-handed semileptonic operators and flavour projectors provides a compact description of the data. The full set of relevant observables is required to impose gauge consistent relations across sectors, constrain the mixing pattern encoded in $\lambda_q$ and prevent improvements in one channel from reappearing as tensions elsewhere. Within this framework, a focused subset of measurements provides the dominant sensitivity to $(C_1,C_3,\beta_q)$. A regression tree algorithm reproduces the non Gaussian shapes seen in the fit and enables efficient uncertainty propagation and correlation studies. Allowing the singlet and triplet coefficients to vary independently, with mixing confined to the second and third quark generations and no lepton mixing, yields regions consistent with current measurements and removes the spurious relation between $R_{D^*}$ and $\mathrm{BR}(B^+\to K^+\nu\bar\nu)$ from previous studies.

\end{document}